\documentclass[useAMS,usenatbib]{mn2e}
\usepackage{xcolor}
\usepackage{amsmath, float}
\usepackage{txfonts}
\usepackage{fleqn}   
\usepackage{graphicx}
\usepackage{epstopdf}
\usepackage{verbatim}
\usepackage{mathtools}
\usepackage{bm}
\usepackage{hyperref}

\title[HD+B model for PWN]
{Introducing the HD+B model for pulsar wind nebulae: \\
a hybrid hydrodynamics/radiative approach}
\author[B. Olmi \& D. F. Torres]{
 B. Olmi$^{1, 2, 3}$ \thanks{barbara.olmi@inaf.it} \& D. F. Torres$^{2, 3, 4}$\thanks{dtorres@ice.csic.es}\\
$^{1}$INAF - Osservatorio Astrofisico di Arcetri, Largo E. Fermi 5,
I-50125 Firenze, Italy\\
$^{2}$Institute of Space Sciences (ICE, CSIC), Campus UAB, Carrer de Magrans s/n, 08193 Barcelona, Spain\\
$^{3}$Institut d'Estudis Espacials de Catalunya (IEEC), 08034 Barcelona, Spain\\
$^{4}$Institucio Catalana de Recerca i Estudis Avanccats (ICREA), Barcelona, Spain}

\begin{document}
 
\date{Accepted / Received }

\maketitle

\label{firstpage}

\begin{abstract}
Identification and characterization of a rapidly increasing number of pulsar wind nebulae is, and will continue to be, a challenge of high-energy gamma-ray astrophysics. 
Given that such systems constitute -by far- the most numerous expected population in the TeV regime, such characterization is important not only to learn about the sources per se from an individual and population perspective, but also to be able to connect them with observations at other frequencies, especially in radio and X-rays. 
Also, we need to remove the emission from nebulae in highly confused regions of the sky for revealing other underlying emitters.
In this paper we present a new approach for theoretical modelling of pulsar wind nebulae: a hybrid hydrodynamic-radiative model able to reproduce morphological features and spectra of the sources, with relatively limited numerical cost.
\end{abstract}


\section{Introduction}
\label{sec:intro}

Pulsar wind nebulae (PWNe) are generated by the interaction of a relativistic, cold, magnetized wind emanating form a pulsar, with the surrounding material.
That interaction induces the formation of a termination shock (TS) at which the pulsar wind is slowed down and heated. 
The nebula, generally shining in a very broad band of energies, from radio to gamma-rays, arises as non-thermal emission from the shocked wind, due to the interaction of the relativistic particles with the nebular magnetic and photon fields, essentially via synchrotron and inverse Compton processes.
Studying these nebulae is critical for understanding the pulsar complex (including the supernova remnant, the pulsar wind nebula, and the pulsar itself), the electrodynamics of the magnetized rotators, how their magnetospheres generate the wind, the acceleration of leptons up to very high energies, their energy distribution, and how the latter feedback on the surrounding interstellar medium. 
However, understanding the time-evolution, spectral properties, and morphology of this kind of systems have proven to be a challenge, and we are yet to see 
a versatile morpho-spectral model of PWNe.
Here, we aim to make a significant step in this direction.

In general, two classes of theoretical models have been developed for PWNe: hydrodynamical (HD) or magneto-hydrodynamical (MHD) models (e.g., \cite{Bucciantini:2002b,vanderSwaluw2004, 
Del-Zanna:2006, Olmi:2013, Olmi:2014,Temim2015,Kolb2017}) and radiative models (e.g., \cite{Venter2007,Zhang2008,Gelfand2009, Bucciantini2011, Tanaka2010,Zhu2018,Martin2012, Torres2014,Martin2016,Torres2018b,Torres2019}). 
The first class provides spatial dependencies, like the distribution of the density, pressure, and PWN magnetic field (if in MHD) as well as the time-evolution of the PWN radius. 
They focus on morphology at different scales and much less (if at all) on radiation being emitted by relativistic particles. 
The latter, instead, is generally computed in detail only from radiative models, where the focus is in the prediction or interpretation of measurable spectral energy distributions. 
These latter models aim at analyzing the impact of time-dependent energy losses on an injected population of particles, but do not consider any morphology and do not produce an energy-dependent map of the nebula. 
Inherent to these radiative models, then, the derived density, magnetic field, and pressure within the nebula have zero gradients. 
It is clear that one model class' caveats are the other's strengths.

Combining magneto-hydrodynamic (MHD) and emission models (e.g., by adjoining or post-processing the latter after the former output) to give time- and spatially-dependent simulations for the PWN evolution would be ideal.
However, if we are thinking of fitting results against observations, 
whatever process is followed in order to achieve this, it has to be versatile, and allow running thousands of different incarnations of model parameters representing different physical situations.
How can this be done? For how long an evolutionary scale?
 
A first attempt to combine particle transport with morphological MHD models was done in \citealt{Porth:2016}.
More recently \cite{vanRensburg2018,vanRensburg2020} are working on introducing a model for 
exploiting morphological data via a spatio-temporal leptonic transport code.
This approach, then, focus on enlarging initially 0D radiative models into at least 1D models, making them able 
to study diffusion processes.
Whereas their recent results are promising and worth of further investigation,
increasing complexities of PWN morphologies, not driven only by diffusion, will hardly be described in such an approach.
Perhaps the most advanced trials into this kind of model combination have been done to date by the Firenze group (see e.g., \citealt{Del-Zanna:2006, Volpi2008,Olmi:2013, Olmi:2014}, from which we inherit experience.  
They have obtained simulated maps for synchrotron and inverse Compton emission with the parameters relevant to a 2D axis-symmetric MHD simulation of Crab (in agreement with its torus/jet structure). 
Their approach was to couple an additional equation to 
the MHD evolution that gives account of the maximal energy of particles. 
Once this is determined, they used it in an a priori assumed particle spectrum and computed the radiation yield. 
However, this approach has not yield to convergence, i.e., the parameters fitting the morphology were not fitting the spectra: for instance, the magnetic field needed in the nebula to reproduce the inner morphology of Crab, was smaller by a factor of $\sim$3 than the one needed from radiative models able to reproduce the spectra.
This mismatch reveals that something in the basis of the approach may need refurbishment. 
For reasons discussed in more detailed below, it is also now well known that 2D MHD models generate unwanted issues in the description of the magnetic field due only to dimensionality.
This makes for unrealistic results of PWNe at regions beyond the termination shock (TS), as well as 
along the symmetry axis of the simulation when only one hemisphere is modelled \citep{Del-Zanna:2006, Olmi:2013, Olmi:2014, Porth:2013, Porth:2014, Olmi:2016}. 

Great progress has been recently achieved in obtaining the first 3D MHD PWN simulations  \citep{Mizuno2011,Porth:2013, Porth:2014, Olmi:2016}.  
In fact, what was known as the sigma problem (how the nebula magnetization evolves with distance from the pulsar, starting from being Poynting-dominated to becoming particle-dominated) has been studied in detail (and largely solved) using these 3D MHD simulations  \citep{Porth:2013, Porth:2014, Olmi:2016}.  
They  allowed for new mechanisms for field dissipation to arise, and observed the appearance of kink instabilities, which are only excited in 3D. 
But as discussed below in more detail too, these simulations are so costly that it is not possible to imagine running them to describe PWNe of more than a thousand years (in a feasible computational time, that is), and less to think on 3D MHD as a way of identifying and characterizing detections that are not yet known otherwise, and counted by the hundreds.
The next generation of gamma-ray telescope, in particular with the forthcoming Cherenkov Telescope Array (CTA), will in fact detect hundreds of new PWNe.
Those sources are also expected to be the dominant component of the gamma-ray sky \citep{de-Ona-Wilhelmi:2013, Klepser:2013, H.E.S.S.-Collaboration:2018a}, making their identification fundamental  not only to gain insight into PWN per se, but also for the removal of this intense component with the aim of revealing background sources that may otherwise be missed.

We are in pursue of a new way of modelling PWN, covering both spectra and morphology in a time-dependent, multi-frequency setting. 
This should be relatively fast and versatile so as to be able to foresee, in the near future, extensions of our approach that would allow a direct 
fitting to observational features. 
Here we propose such a new theoretical approach, which we dubbed the HD+B model.
It will be based on performing 2D HD simulations, and attaching to it a magnetic field based on reasonable physical assumptions.
We introduce all details next, and provide the first application with the young, composite, PWN G21.5-0.9. 
%

\section{The HD+B method}
\label{sec:HD+B}

\subsection{Why 2D HD? Why not 3D MHD or 2D MHD?}

Whereas it is clear that a lower dimensionality (i.e., 1D) would not contain any morphological information, we here consider why a higher 
one (3D), or the use of 2D but in the MHD setting are not appropriate for our aims.

In the last years, PWNe have been successfully modelled with 3D relativistic MHD simulations, showing impressive results when accounting for their full dynamical structure, in particular being finally able to reproduce the complex structure of the magnetic field \citep{Porth:2014, Olmi:2016, Olmi:2019, Olmi:2019a}. 
However, 3D simulations were also shown to require a huge amount of numerical resources and computational time to reproduce only a limited part of the evolution of a PWN.
Millions of cpu hours and months of simulation runs are currently needed to obtain just $\sim1/5$ of the Crab nebula age.
Despite a 3D approach is the most advanced tool to model a single PWN --particularly if it is well characterized otherwise--, those simulations are obviously impossible to implement for numerous systems, the very most of them being much older than the Crab nebula.
For instance, identifying first, and characterizing later the several hundreds PWNe that will be discovered just by the Cherenkov Telescope Array using 3D simulations is not the right approach:
Running simulations as in  \cite{Porth:2013,Porth:2014,Olmi:2016} for $\sim$50000 years, several parsecs in size, for a large phase space of PWNe, is simply impossible with the current computational reach.

The more natural step back one may think of could then be using 2D MHD simulations instead. 
But the 2D approximation in the MHD scheme was shown to introduce strong artifacts to the dynamical evolution of the system, making 2D models able to account for the properties of the very inner nebula only, in the vicinity of the termination shock \citep{Del-Zanna:2006, Volpi2008, Camus:2009, Olmi:2014, Olmi:2015}. 
In such 2D MHD simulations, the magnetic field appears in fact to be strongly squeezed around the polar axis, with a complete loss of its real geometry. 
The reason behind this is well understood, the inability of the magnetic field to dissipate in the third direction of space produces an unrealistic 
scenario, it does not activate instabilities as in a 3D setting, hampering the simulation reliability. 
In fact 3D instabilities, as the kink one, produce not only a strong channel for magnetic dissipation, allowing to increase the initial magnetization of the pulsar wind up to values greater than unity at injection, but also break the toroidal geometry of the field, producing a poloidal component immediately outside of the inner nebula, that becomes dominant at the boundary. 
This on the contrary does not happen in 2D, where the toroidal shape of the magnetic field is strongly conserved.
In the vicinity of the termination shock (TS), where also in 3D the deformation from the toroidal geometry is very low, the magnetic field is quite well represented, and this is the reason why 2D models are still able to account for the properties of the inner nebula.
On the contrary, when moving away from the TS, this artificial deformation makes the system very different from what observed in 3D.
For the former reasons, even paying the price of loosing direct information about the field, an approach based on HD instead of MHD makes 2D simulations more reliable, and the obvious choice if the focus is put in the overall morphology of the nebula rather than in the intricacies of its innermost part.

%
Let us analyze the situation from a different perspective.
According to theoretical models of pulsar magnetospheres as well as the observations of pulsations across all frequencies, the pulsar wind is expected to be highly magnetized near the pulsar light cylinder.
There, $\sigma$, the ratio between the
Poynting and the particle kinetic energy fluxes is large, $\sigma \gg 1$ (see e.g., \cite{Arons2012} and references therein).
But already from the first models of the Crab PWN, it was noted that its nebula should be particle dominated \citep{Rees1974,Kennel1984,Begelman1992}, what was confirmed in full radiative modelling of other PWNe as well \citep{Tanaka2010,Bucciantini2011,Torres2013,Torres2014,Zhu2018}, where the average $\sigma \sim 10^{-3}$. 
The change in magnetization is so severe that cannot be accounted by dissipation along the flow, see, e.g.,  \cite{Lyubarsky2009,Lyubarsky2010}
nor by magnetic dissipation 
from reconnection in stripes of pulsar wind with different polarities  \citep{Coroniti1990,Lyubarsky2001}.  
Significant advancement came recently, following earlier suggestions that the dimensionality of the simulations may hamper the dissipation \citep{Begelman1998}.
Indeed, the 3DMHD numerical simulations of the Crab PWN \citep{Mizuno2011,Porth:2013,Olmi:2016} showed that 
the onset of kink instabilities leads to a large amount of dissipation with respect to that observed at lower dimensionality.
Such strong dissipation renders the total pressure nearly constant across the whole PWN. Away from the TS, 
also the magnetic field variation is roughly constrained to a constant, within a factor of a few. 
These facts are likely the reason behind the great success of radiative models based on a single zone for reproducing the spectral energy distribution of PWNe, a perspective first advanced by \cite{Gelfand2017}.
We here note that this applies as well to the 2D HD approach, which is also producing a more uniform pressure distribution than what would result from a 2D MHD simulation, and thus is closer to reality overall.
We show this is in more detail next.

\subsection{Combining 2D HD with radiation}

We then propose a novel approach based on the use of 2D HD simulations combined with radiation, with the magnetic field included thanks to dedicated numerical tracers. 
Details follow.

\subsubsection{Details on the numerical tool and setup}
\label{subsec:details_setup}

The simulations used in this paper have been performed with the PLUTO numerical code \citep{Mignone:2007a}, using the relativistic HD module and Adaptive Mesh Refinement (AMR) \citep{Mignone:2012}. 
AMR is chosen to speed up the computation maintaining the necessary high level of resolution near the grid center, where the pulsar wind is injected and where the TS must be properly resolved to ensure the correct evolution of the system. 

Without AMR, in the general situation, if we were to properly characterize the TS and the wind injection region, we would be expending an unwarranted 
amount of computational time in the outer parts of the PWN. 
Logarithmic-spaced scales or even nested grids can also be used to increase resolution at the grid center, but even in those cases AMR is the optimal solution to speed up computation, being sure to increase resolution only where it is really needed (see a clear example in Fig.~\ref{fig:vort_res} of Appendix~\ref{sec:a1}).

We define a base grid of $272\times544$ cells and four AMR levels, to get an equivalent maximum resolution of $[4352]\times[8704]$ cells at the finest level.
With AMR the resolution is automatically increased near shock fronts. 
We also impose a spatially dependent condition to get the maximum resolution only near the grid center, where the pulsar wind is injected from a reset region.
The setup of the base grid and AMR levels has been determined as the compromise between a sufficient resolution in the nebula and the lowest numerical cost of each run.
When defining the resolution we had also considered that in the inner region it must be high enough to ensure a good mixing of the material injected into the nebula, meaning that the turbulent vortexes must develop on dimensions smaller than the nebula radius. 
This ensures a correct mixing of the freshly injected material with the older one in the surroundings.
On the other hand, the resolution must not be too high at the contact discontinuity, where the Rayleigh-Taylor instability (RT) mixes the PWN material with the one from the ejecta. 
The efficiency of this mixing process was shown to be resolution dependent, and in HD, without the support provided by the magnetic pressure, the RT fingers may penetrate the nebula bubble very deeply \citep{Blondin:2001}.
This effect was not noticed in MHD simulations \citep{Porth:2016}, where the presence of the magnetic field seems to act as a stabilizer and the fingers only penetrate a fraction of the total volume, in agreement with observations. Some additional discussion regarding this issue can be found in Appendix~\ref{sec:a1}.

The stability of the code is augmented thanks to the definition of a time-dependent radius of the wind injection region, that increases with time in order to always be in the vicinity of the TS. This helps removing eventual artifacts produced near the grid axis due to axisymmetry.
We use a second order Runge-Kutta time integrator and the Harten-Lax-van Leer (HLL) Riemann solver  for discontinuities.  

In the HD scheme, the dynamical variables evolved by the code, consistently with the relativistic hydrodynamic equations, are 
density, pressure, and velocity $(\rho,\, P,\, \textbf{v})$.
A few passive numerical tracers, initialized within the TS and then advected in the nebula after solving the jump conditions at the shock, have been added to the code.
A first tracer is used to firmly identify the material belonging to the PWN from the one coming from the outer medium. 
Since numerical mixing tends to melt the fingers coming from outside with the PWN material, that tracer allows us to define at each time step the PWN and it is used both for defining the radius of the nebula and for retracing \emph{a posteriori} the age of the injected particles.
Other tracers have been defined in order for the magnetic field to be included as a non-dynamical variable; in particular we trace the maximum and break energies of the emitting particles, subjected to synchrotron and adiabatic losses, and the PWN material.
The exact definition of how those tracers are treated can be found in the following Section~\ref{subs:HD+B}.
%

\subsubsection{ Details on the physical model of the PWN}
	
Following previous literature  (see for example \citealt{Olmi:2014} for a similar approach in the RMHD case) we generate the PWN by continuously injecting a pulsar wind at the center of the SNR shell, characterized by the spin-down luminosity
\begin{equation}\label{eq:Lt}
	L(t)=L_0/(1+t/\tau_0)^{\beta}\,,
\end{equation}
where $L_0$ is the initial luminosity, $\tau_0$ the spin-down time and $\beta=(n+1)/(n-1)$ is connected to the pulsar braking index $n$, herein considered to be $n=3$, leading to $\beta=2$.
The adopted spin-down dependence in time is consistent with assumptions taken in radiative models.
The pulsar wind is then characterized by a density 
\begin{equation}\label{eq:rho}
	\rho(r,\, t)=\rho_0(t)/(r/r_0)^{2}\,,
\end{equation}
where $\rho_0(t) = L(t)/(4\upi \gamma_w^2 c^3 r_0^2)$ and $r_0=1$ ly is the reference spatial dimension.
The pressure is given by the solution of the ideal equation of state for a relativistic plasma (with adiabatic index $\Gamma=4/3$), imposing the ratio at injection $p_0/\rho_0\sim 10^{-2}$ for representing the cold pulsar wind.
The simulation is initialized with a PWN of age $t_\mathrm{ini}=80$ yr. 
Values at $t=t_\mathrm{ini}$ are selected according to observational constraints, whenever possible, or are obtained from radiative only models, e.g., \cite{Torres2014}. 
The system is then evolved up to its actual age ($t_\mathrm{age}$).
The initial bulk Lorentz factor of the wind is imposed to be $\gamma_w=10$, even if the expected value is much higher ($10^4 \lesssim \gamma_w \lesssim 10^6$, \citealt{Kennel:1984a, Bucciantini:2003}).
Simulations cannot manage such high values of the Lorentz factor. 
However, this numerical limitation was shown not to affect the dynamics, being correctly reproduced since the relativistic nature of the wind is already ensured \citep{Del-Zanna:2006}.

The cold supernova ejecta are modeled following the usual prescription (e.g. \citealt{van-der-Swaluw:2001, Del-Zanna:2004}): for $r\leq R_\mathrm{ej}(t_0)=v_\mathrm{ej}t_0$, with $R_\mathrm{ej}(t_0)$ the outer radius of the ejecta shell at time $t_0$, the ejecta are characterized by a (high) constant density and a Hubble-type profile for the velocity 
\begin{equation}\label{eq:ejecta_prof}
 \rho_\mathrm{ej}(t_0) = \frac{3}{4\upi} \frac{M_\mathrm{ej}}{R_\mathrm{ej}^3(t_0)}\,,
\quad v(r,t_0) = v_\mathrm{ej} \frac{r}{r_\mathrm{ej}(t_0)}\,, 
\quad v_\mathrm{ej}=\sqrt{\frac{10}{3}  \frac{E_\mathrm{SN}}{M_\mathrm{ej} }} \,, 
\end{equation}
with $E_\mathrm{SN}$ the supernova explosion energy and $M_\mathrm{ej}$ the ejected mass.
The ejecta expands into the surrounding interstellar medium (ISM), with low density ($\lesssim 1$ particle/cm$^3$, $T\sim 10^4$ K and fully ionized).

Initial values of the parameters defining the surrounding medium and the progenitor star will depend on the particular object, and will be discussed in this paper in connection with our first application presented in Section~\ref{sec:appG21}. 
Usually one or more or those parameters will not be well constrained, and different values of the mass of the ejecta or the ISM density may, in principle, lead to a different evolution of the PWN radius with time. 
We aim at investigating the degeneracy of the model prediction when varying considerably some of the initial parameters in a future
work.

\subsubsection{Coupling the HD model with B}
\label{subs:HD+B}

The HD simulation is coupled to the radiative model with an energy-based prescription for the field, discussed in the following.
Being a passive tracer, the magnetic field is kept disentangled from the dynamics, to avoid the artificial modifications observed in 2D MHD models.
We consider that the total energy stored in the PWN ($E_\mathrm{PWN}$), which is coming from the energy released in the system by the pulsar, can be divided in two components: 
the energy that goes into particles, $E_P$ is (following e.g. \citealt{Torres2014,Martin2016})
\begin{equation}\label{eq:E_P}
	E_P R_\mathrm{PWN} = (1-\eta )\int^{t}_{0} L(t^\prime)R_\mathrm{PWN}(t^\prime) dt^\prime \,,
\end{equation}
where $\eta = L_B(t)/L(t)$ is the magnetic energy fraction, with $L_B(t)$ the magnetic power; and the energy that goes into the magnetic field $E_B = V_\mathrm{PWN}(t) B^2(t)/(8\upi)$
\begin{equation}\label{eq:EB}
	E_B R_\mathrm{PWN} = \eta \int^{t}_{0} L(t^\prime)R_\mathrm{PWN}(t^\prime) dt^\prime \,,
\end{equation}
where $V_\mathrm{PWN}(t)=[4\upi R_\mathrm{PWN}^3(t)]/3$ is the volume of the PWN at time $t$.
It can be noticed that Eq.~\ref{eq:EB} is actually equivalent to
\begin{equation}
	\frac{d E_B}{d t} = \eta L - E_B \left( \frac{d R_\mathrm{PWN}}{d t} \frac{1}{R_\mathrm{PWN}}\right)\,,
\end{equation}\label{eq:dEBdt}
and it takes into account properly the adiabatic losses.
The magnetic field can be then obtained from the magnetic energy as 
\begin{equation}\label{eq:Bt}
	B(t) = \sqrt{8 \upi \frac{E_\mathrm{B}(t)}{V_\mathrm{PWN}(t)}} =\sqrt{ 8\upi  \eta\frac{  E_\mathrm{PWN}(t)}{V_\mathrm{PWN}(t)}}\,.
\end{equation}
We use Eq. (\ref{eq:Bt}) to introduce the magnetic field tracer in our description, where the energy content of the PWN at each time step is naturally computed from the numerical simulation, recalling that we are in a fully hydrodynamical scheme and thus: 
\begin{equation}\label{eq:Epwn}
E_\mathrm{PWN}(t)=\left[1/(\Gamma-1)\right] PV_\mathrm{PWN}(t).
\end{equation}

In this scenario the magnetic field has no back-reaction on the dynamics,
a simplifying assumption that is considered acceptable for particle-dominated nebulae. 

The field, however,  is linked to the physical evolution of the system, as ensured by the hydrodynamical tracking of the pressure. 
In practice, the magnetic field is modelled as an ultra-relativistic gas, and any energy transfer between the particle and magnetic field components is neglected.
This allows us to correctly account for the particles losses due to synchrotron radiation in addition to adiabatic expansion. 
In fact, we now have a magnetic field which is both, spatially- and time-dependent, and obtained 
from the evolution and distribution in space of the total pressure.
This field can be used in the general expression for the particles losses. 
These are taken into account using the same approach as discussed in \citet{Del-Zanna:2006}. 
Integrating along the streamlines the equation for the time evolution of the energy of a single particle in the post shock flow, $\epsilon$, and combining it with the conservation equation for the mass, $(\partial \rho/\partial t) + \mathbf{v} \cdot \mathbf{\nabla} \rho=0$, leads to the following equation for the maximum energy of the emitting particles ($\epsilon_\mathrm{max}$)
\begin{equation}\label{emax_dz}
	\frac{\partial}{\partial t} \left( \gamma_w \rho^{2/3} \epsilon_\mathrm{max}\right) 
	+ \mathbf{\nabla} \cdot \left( \gamma_w \rho^{2/3} \epsilon_\mathrm{max} \mathbf{v} \right) 
	=  - \left( \frac{4 e^4}{9 m^3 c^5} \frac{B^2}{\gamma_w^2}\epsilon_\mathrm{max} \right) \rho^{2/3}\,.
\end{equation}
Namely $\epsilon_\mathrm{max}$ considers the integrated synchrotron and adiabatic losses, being thus the remaining energy for a particle that was injected into the system with an initial energy $\epsilon_0$.
The maximum energy must be initialized to a value that ranges between the expected maximum energy (from the high energy spectral cut-off) and the pulsar voltage \citep{Bandiera:2008}, so that $10^9 \lesssim \epsilon_\mathrm{0} \lesssim e /(2 m_e c^2)\times (3\dot{E}/2c)^{1/2}$.
Finally, a similar approach is used to evolve the tracer of the break energy, $\epsilon_b$, initialized with a fixed value of the order $10^5 - 10^6$.

\subsubsection{Properties of the emitting particles}

We define the particle distribution function as 
 \begin{equation}\label{eq:f}
	f(\epsilon) = K_0 \left( \frac{\epsilon}{\epsilon_b}\right)^{-p_i}  \exp{\left(-\frac{t_{\mathrm{particle}}}{\tau_\mathrm{sync}(t_\mathrm{age})}\right)}, \hspace{.5cm} {\rm for} 
	  \quad 0 \leq  \epsilon \leq \epsilon_\mathrm{max} \,,
\end{equation}
where $\epsilon=E/(m_e c^2)$ is the particle Lorentz factor. 
The exponential factor in the previous equation takes into account the appropriate distribution of particles in energy, according to their actual age ($t_{\mathrm{particle}} < t_\mathrm{age}$) compared to the synchrotron life time at the present age of the system, with $\tau_\mathrm{sync}=m_e c^2 /[(4/3)\gamma_w  \sigma_T c (B^2/8\upi)]$.
Thus, the latter is not fixed a priori but it comes from the consistent evolution of the dedicated magnetic field tracer in the simulation, having its corresponding spatial distribution and time evolution.

As it has been usually assumed in radiative models of PWNe, the subscript $i=\{L,\,H\}$ indicates two different populations of emitting particles: the particles responsible for the low-energy emission,  where $\epsilon \leq \epsilon_b$, and the ones responsible for the high-energy ones, where $ \epsilon_b < \epsilon \leq \epsilon_\mathrm{max}$.
Usually, in radiative models the normalization constant of the spectrum, $K_0$, is simply obtained by fitting the spectral data.
In e.g., \citet{Del-Zanna:2006, Volpi2008} the normalization has been defined differently for the two particles families.
Here we try to reduce as much as possible the free parameters of the fit, and we adopt the 
most natural way to normalize the spectrum: we link $K_0$ to the thermal energy by equating
\begin{equation}\label{eq:sp_norm}
	\frac{3\mathcal{P}}{( m_e c^2)} =  \int_0^{\epsilon_\mathrm{max}} f(\epsilon)\epsilon d\epsilon\,,
\end{equation}
where $\mathcal{P}$ is the thermal pressure. 
Recalling that the simulation is not considering the magnetic pressure, this $\mathcal{P} < P$ (the pressure obtained from the hydrodynamical simulation directly),
since the simulation implicitly assumes that all the energy injected by the pulsar goes into the thermal component, which is not the case.
In reality, part of the energy injected by the pulsar must go into magnetic energy (or pressure) during the evolution, since the real system has not $B=0$.
Since we are injecting the real power (as defined, for instance by true spin-down luminosity) from the pulsar into the system, we have to account for this difference, otherwise we will be overestimating the amount of energy stored in particles.
The simulation-obtained thermal pressure $P$ must be then rescaled by a factor $c_1$ taking this effect into account. 
That correction factor is fixed in our approach by computing the expected thermal and magnetic pressures at $t_\mathrm{age}$ as
\begin{equation}\label{eq:corr1}
	c_1 = 1 - \int_0^{t_\mathrm{age}} P_B(t^\prime) dt^\prime  
		\left[ \int_0^{t_\mathrm{age}}  P(t^\prime) dt^\prime \right]^{-1} \,,
\end{equation}
where with $B(t)$ is given by Eq.~\ref{eq:Bt}. 
So, we are actually extracting from the simulated, integrated-in-time pressure, the part that in reality must have gone to powering the field.

Moreover, the HD model is also not taking into account the radiation losses experienced by the system, since particles are not included in the dynamics and radiation has no back-reaction on the system during its evolution. 
Radiation our model is in fact computed \emph{a posteriori} of the hydro simulation, 
considering that the emitting particles are continuously injected at the TS with the distribution function defined  in Eq.~\ref{eq:f}.
A consistent evolution of the particles, and its coupling with the dynamics, would only be possible with an additional coupling with PIC-like codes, which cannot be used in this case due to the very broad range of spatial (and time) scales involved (see for example \citealt{Sironi:2009}).
If we were not to correct by the radiation that particles emit,  the entire number of particles injected in the system from 0 to $t_\mathrm{age}$ would still be available at $t_\mathrm{age}$, overestimating the real number of particles since a part of them would have been --in reality-- lost due to radiation processes. 
Thus, the pressure $P$ must then be corrected by a second factor, $c_2$, accounting for the energy lost in radiation from the nebula's birth to $t_\mathrm{age}$. 
This too can be formally obtained from our simulations, by computing in the post-processing the following factor
\begin{equation}\label{eq:corr2}
		c_2 = \frac{ \int_0^{t_\mathrm{age}} \int_0^{\epsilon_\mathrm{max} }  \left[ \epsilon/ \epsilon_b \right]^{-p_i} d\epsilon dt^\prime  }  
		         { \int_0^{t_\mathrm{age}} \int_0^{\epsilon_\mathrm{max}}   \left[ \epsilon/ \epsilon_b \right]^{-p_i} e^{-t^\prime/\tau_\mathrm{sync}(t^\prime)} d\epsilon dt^\prime}  \,.
\end{equation}
The numerator of the previous formula gives the total number of emitting particles injected into the system during its history, from its birth up to $t_\mathrm{age}$. 
On the contrary, the denominator takes into account that part of the total injected particles must have been lost via synchrotron radiation, i.e. their lifetime is weighted with the synchrotron age that particles have at each time-step of the evolution. 
The ratio of the two gives an indication of the amount of particles that are still available at the present age.
We stress again that these corrections, $c_1$ and $c_2$, are computed in the post-processing of the dynamical simulations, using the entire set of time-dependent data that resulted from it.

The overall normalization of the spectrum is then obtained from physical arguments. 
Taking all of this into account, the expected thermal pressure is 
\begin{equation}
\mathcal{P}=c_1 c_2 P ,
\end{equation}
 and this is what we use in Eq. (\ref{eq:sp_norm})
to normalize the number of particles, with $c_1$ and $c_2$ numerically determined from the simulation.

\section{Application to the G21.5-0.9 PWN}
\label{sec:appG21}

\subsection{Size and field today}

%
G21.5-0.5 is a young PWN, located at the center of a composite supernova remnant. 
It has an estimated age of $t_\mathrm{age}=870^{+200}_{-150} $yr \citep{Bietenholz:2008} and located at a distance of  $d=4.7$ kpc  \citep{Camilo:2006, Tian:2008}.
The PWN shows a spatial extent with a radius of $\sim 3$ ly and the puzzling feature of having almost the same size at radio and X-ray frequencies, possibly indicating the dominance of diffusion processes near the PWN boundary, where thermal filaments are located, as seen at higher energies.
 
\begin{figure}
	\includegraphics[width=\columnwidth]{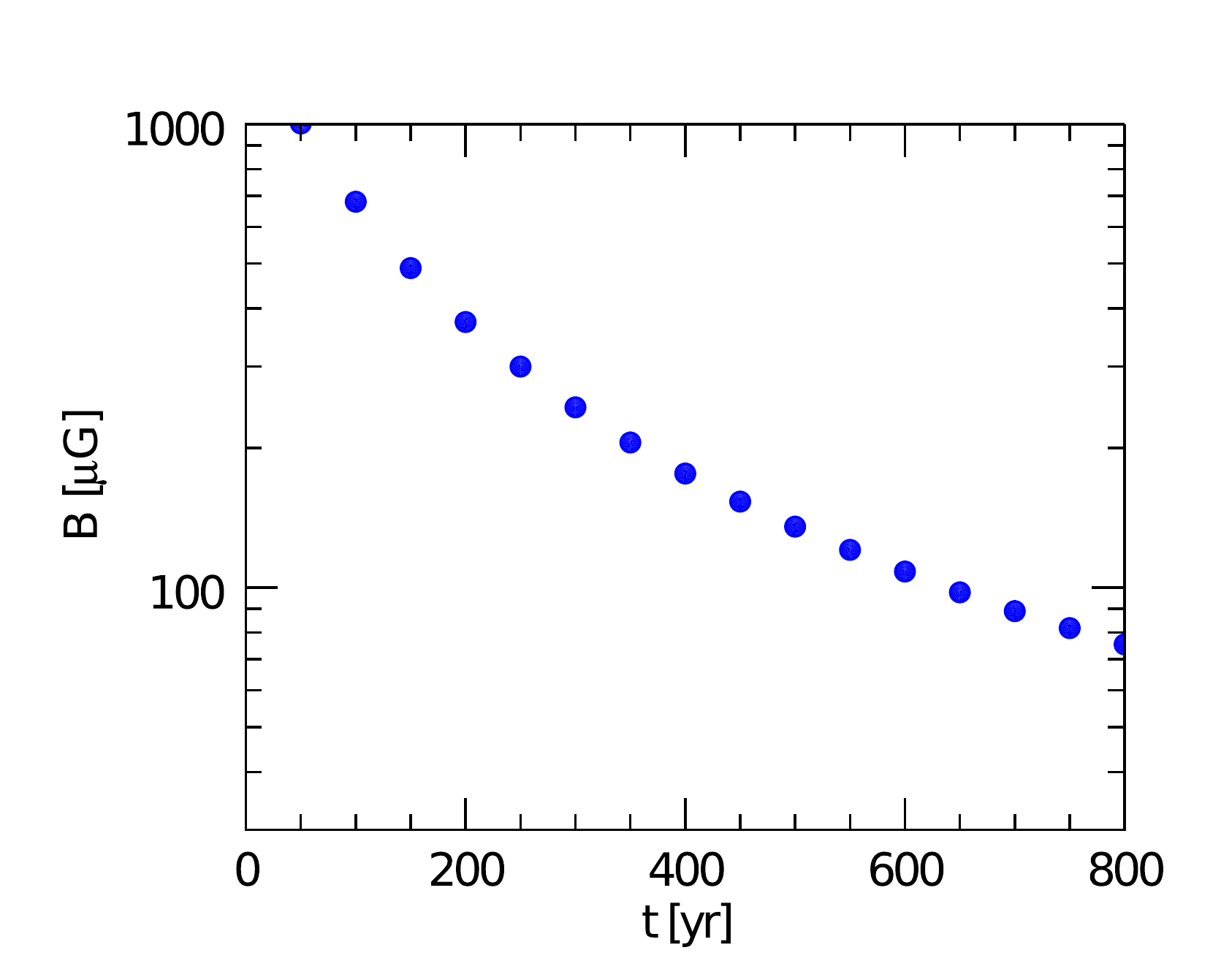}
	\caption{Evolution of the volume-averaged value of the magnetic field tracer (in $\mu$G) with the simulation time (in yr). 
	At the actual age of the system, when the PWN radius matches the expected dimension of $R_\mathrm{PWN}\simeq 3$ ly, the mean field is of order of $82\mu$G. }
	\label{fig:Bave}
\end{figure}

The simulation is initialized with the values: $L_0 = 5 \times 10^{37}$ erg/s for the initial luminosity and $\tau_0=3985$ yr for the spin-down time, as deduced from observational constraints \citep{Matheson:2010}.
No information are available for the progenitor star and the supernova shell, so that we use the values obtained before 
from radiative models: $E_\mathrm{SN} = 10^{51}$ erg and $M_\mathrm{ej} = 8 M_\odot$ \citep{Torres2014}. 
The ISM is modeled as a low density medium with 0.1 particles/cm$^3$, low pressure and zero velocity.
Radiative models predict an average value of the magnetic field at $t_\mathrm{age}$ of $\sim 70\,\mu$G, with an initial magnetic fraction of $\eta = 0.04$.
That value of the magnetic fraction does not appear to be able to reproduce morphology and spectral features, 
and we have increased it by a factor of 2 in a successive run, leading to a slightly higher magnetic field in the nebula at the end (see Fig.~\ref{fig:Bave}). Such differences in the value of magnetic fraction needed are reasonably expected, given the different methodological approach. In both cases, however, the PWN has an instantaneous distribution of energy that is heavily particle-dominated. 

The system is evolved in order to match the expected dimension of of the nebula, up to a final age of the simulation of $t_s=750$ yr. 
We then estimate an age of the system of  $t=t_\mathrm{ini} + t_\mathrm{run}=830$ yr, fully compatible with the one deduced from observations.
Starting with an initial field of a few mG, the average field in the nebula at the final age is of $82\mu$G.
The difference with the radiative model estimation is not that surprising, since pure radiative models do not account for the dynamics and a small difference might be naturally expected.

\subsection{Comparing HD+B with MHD}
\label{subs:cfrBHD-MHD}

\begin{figure*}
        \includegraphics[width=14cm]{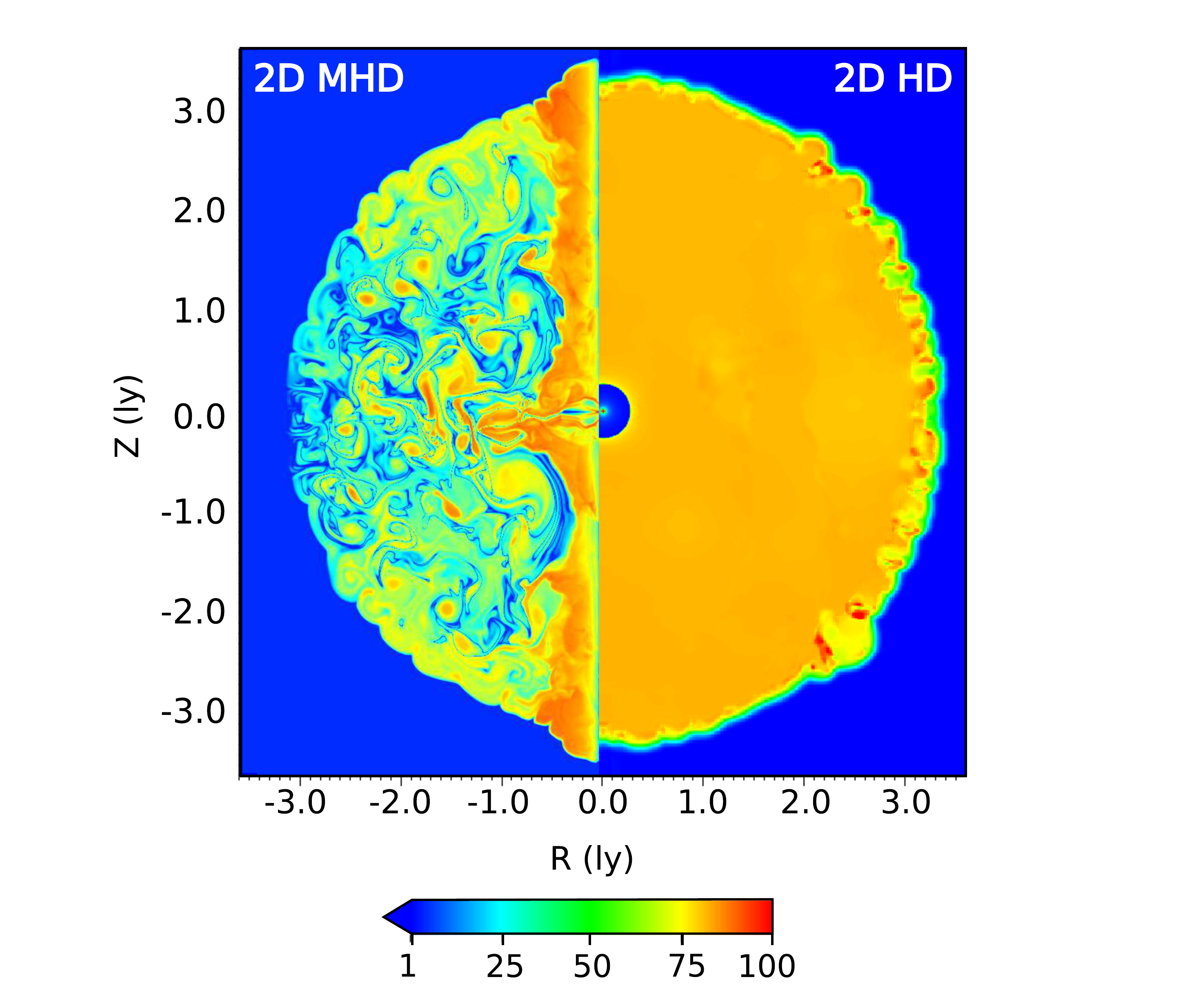}\;
	\caption{Comparison of the 2D map of the magnetic field intensity in the nebula under different models: on the left-hand side the magnetic field arising from a MHD simulation of G21.5-0.9; on the right-hand side the magnetic field modelled in the HD+B approach from HD simulations, as detailed in the text. The intensity is given in $\mu$G units and have the same color scale. The ambient magnetic field is zero and the minimum value in color map is saturated to unity just to highlight the nebular field.
	The inner circular structure visible in the right-hand panel is the pulsar wind termination shock, which is spherical due to the assumption of an isotropic wind.}
	\label{fig:B_CFR}
\end{figure*}

In Fig.~\ref{fig:B_CFR} we show an intensity map of the magnetic field, comparing results obtained from an 2D MHD and an HD+B simulation at at the same age of $750$ yrs. 
Both simulations are initialized with the same values, appropriate for modelling G21.5-0.9, as discussed previously. 

As can be noticed if comparing the geometry of the termination shock in the two cases, we have considered an isotropic energy injection (spherical TS)  for our HD models.
However, the usual assumption would be an anisotropic distribution, with the level of anisotropy governed by a free parameter, $\alpha$, that modulates the angular distribution of the energy flux: $F(\theta)\propto (1+\alpha \sin^2\theta)$ with $\theta$ being the colatitude. 
Since we were not going to compare with high-resolution images, for which a correct description of the inner structure of the nebula is necessary, and we have no a priori idea of the level of anisotropy of the wind either (G21.5-0.9 shows a quasi uniform X-ray emission, extending up to the PWN dimension), we prefer to maintain
 the description  as simple as possible, setting the anisotropy parameter to $\alpha=0$. 
Of course, this assumption can be easily relaxed to model different sources if knowledge about the wind anisotropy is achieved.
This figure also illustrates the issues considered in Section 2. 
The artificial deformation of the magnetic field and the nebula in the 2D MHD model can be easily noticed, due to the reduced dimensionality (left-hand panel of Fig.~\ref{fig:B_CFR}).
A strong compression is evident around the polar axis, leading to an artificially confined field  \citep{Olmi:2014, Porth:2014, Olmi:2016}.
Moreover, the lack of an efficient mixing and dissipation of the field into the nebula leads to an average field which is well below the HD+B (or radiative-only) model value, with the latter reached only in the proximity of the polar axis whereas the bulk of the nebula has a very low field  ($\sim 25 \mu$G).
The polar compression of the magnetic field also reflects in the rupture of the spherical shape of the bubble, with some extruding material along the polar axis, which will form polar jets of escaping density if the system is let to evolve longer.

\begin{figure*}
        \includegraphics[width=14cm]{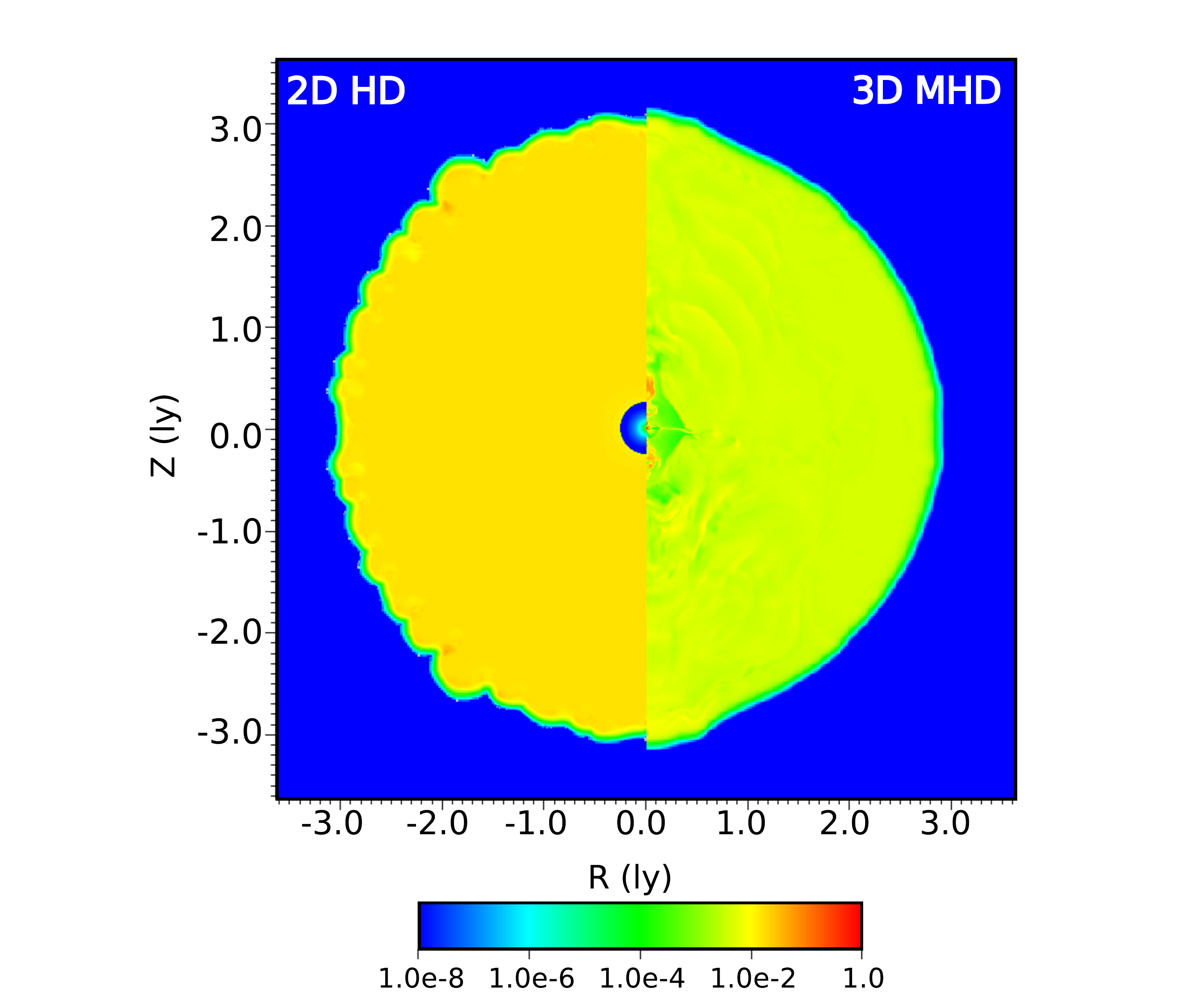}\;
	\caption{
	Comparison of the 2D map of the total pressure in the nebula ($P_\mathrm{TOT} = P_\mathrm{TH} + P_B$) between the 2D HD case (left-hand panel) and the 3D MHD one (right-hand panel). The 3D simulation is the one presented in \citet{Olmi:2016} for the Crab nebula evolved for 250 years. The 2D HD one is produce with the same setup of the one presented in this paper, with parameters compatible with the ones of the 3D simulation, and evolved up to reach a comparable age.
	The maps are in logarithmic scale and normalized to the maximum, for an easier comparison. 
	}
	\label{fig:2dhd_vs_3dmhd}
\end{figure*}

In our HD+B approach, the magnetic field results quite uniform in the whole nebula and with a reasonable average value, given its direct link to the pressure distribution (as can be seen in the right-hand panel of Fig.~\ref{fig:B_CFR}). 
Since the magnetic evolution is not affecting the dynamics, in this case the PWN shell is not artificially deformed.
Comparing with the expected, almost uniform, large-scale distribution of the magnetic field from 3D simulations (see \citealt{Porth:2014, Olmi:2016}), it is evident that the one obtained in the HD+B approach is more representative than the distorted one from 2D-MHD simulations, emphasizing the commentary of Section 2.
This comparison is done explicitly in Fig.~\ref{fig:2dhd_vs_3dmhd}. In there, we show the distribution of the total pressure in the nebula in the 2D HD case (panel on the left) and in the 3D MHD one (panel on the right). 
The 3D simulation is the one of the Crab nebula presented in \citet{Olmi:2016}, with the system evolved for 250 yrs.
The HD simulation is then made ad hoc in order to compare with the 3D one, setting up a run with the same geometrical configuration of the one used in this paper for G21.5-0.9 but changing the parameters of the source to match those used in the 3D simulation.
Of course the two show a complete different structure of the TS that of course reflects in the modification in the inner structure of the nebula. 
This is moreover shaped strongly by the presence of the toroidal field.
But in the sake of a qualitative comparison, the bulk distribution of the pressure in the nebula, when moving outside the very inner part, appears quite similar, with the pressure being almost uniform.
The reason why the HD simulation is more suitable in reproducing the large scales structures of the PWN is evident if comparing the 3D MHD distribution of the pressure with the magnetic field from 2D MHD shown in Fig.\ref{fig:B_CFR}.

We keep in mind, however, that the real properties of the magnetic field must take into account also local variations.
%
%
Our approach is over-simplifying  the geometry of the field, making it unable to reproduce small-scale features, especially those observed in the inner nebula connected to the field morphology.
This kind of complete modelling can be only reached with 3D simulations, albeit, as already mentioned, are impossible to use for a study of numerous evolved systems due to their huge computational and time costs.
From a practical point of view, we also note that such inner nebula fine details will be beyond the resolution of PWN finding factories (like the CTA in gamma-rays) for years to come, so that characterization of PWNe will have to be done at larger scales.

\subsection{Computing the multi-wavelength emission properties in the HD+B approach}
\label{subs:emission}

\subsubsection{Processes considered}

Multi-wavelength emission properties can be determined starting from the simulated nebula, with the association of the particle distribution function,  by computing synchrotron and inverse Compton emission.
The synchrotron spectral power $\mathcal{P}_\nu$ emitted by a particle can be written in the monochromatic approximation as
 \begin{equation}\label{eq:syncP}
	\mathcal{P}^\mathrm{SYNC}_\nu (\nu,\epsilon) = 2\sigma_T c P_B \epsilon^2 \delta(\nu-\nu_m) \,,
\end{equation}
where $P_B=B_\perp^2/(8\upi)$ is the magnetic energy density associated to the local magnetic field component orthogonal to the particle velocity, $\sigma_T$ is the Thomson cross-section, $\nu$ is the observed frequency and $\nu_m=0.29 [3e/(4\upi m_e c)B_\perp \epsilon^2]$ is the maximum emission frequency (see for example \citealt{Rybicki:1979}).
In the present case the component of the field orthogonal to the velocity is simply given as the one resulting from an isotropically distributed field, so that $B_\perp = \sqrt{2/3}B/\gamma_w$ , where the $\gamma_w$ comes from the conversion between the local and observer's reference frame \citep{Del-Zanna:2006}.

%
\begin{figure*}
        \includegraphics[width=1.0\textwidth]{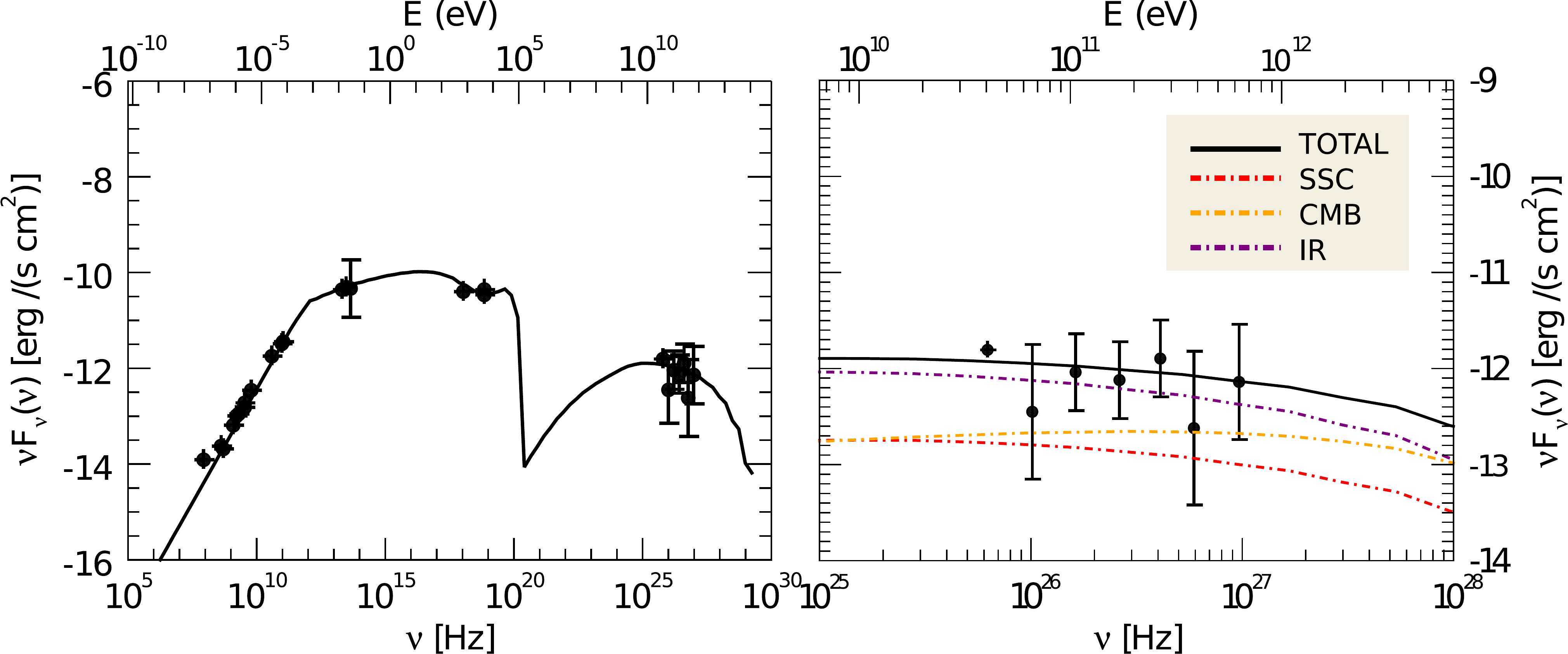}
	\caption{\emph{Left panel}: Integrated total spectrum of the G21.5-0.9 PWN. The black line shows the total contribution of the different emission mechanisms (from synchrotron to IC) to the spectrum, considering all the emitting families and the different photon targets. 
	\emph{Panel on the right}: highlight of the IC emission, with different colors representing the different contributions (as stated by the legend).
	 The underlying observational data in both panels are as in \citet{Torres2014}, to which we refer for the complete list of the observational works considered.}
	 
	\label{fig:spectrum}
\end{figure*}

The inverse Compton (IC) spectral power emitted by a single particle is given by
 \begin{equation}\label{eq:ICP}
	\mathcal{P}^\mathrm{IC}_\nu (\nu,\epsilon) = c h \nu \int \left(\frac{d\sigma}{d\nu^\prime}\right)_{\mathrm{IC}} (\nu^\prime,\,\nu,\, \epsilon) \,n_\nu(\nu^\prime) \,d\nu^\prime \,,
\end{equation}
where $h$ is the Plank constant and $\left(d\sigma/d\nu^\prime\right)_{\mathrm{IC}}$ is the general differential cross-section for IC scattering per unit frequency, accounting for both the non relativistic regime (Thomson scattering) and the relativistic one (Klein-Nishina), which can be expressed as \citep{Jones:1968, Blumenthal:1970}
 \begin{equation}\label{eq:IC_sigma}
	\left(\frac{d\sigma}{d\nu^\prime}\right)_{\mathrm{IC}} = \frac{3}{4} \frac{\sigma_T}{\epsilon^2 \nu^\prime} \left[  2q\ln q + (1-q) \left( 1+2q+\frac{1}{2}\frac{(\Gamma_e q)^2}{1+\Gamma_e q}  \right)   \right] \,,
\end{equation}
with $q=h\nu/(m_ec^2 \Gamma_e) [\epsilon- h\nu/(m_ec^2)]^{-1}$ and $\Gamma_e =4\epsilon h \nu^\prime /(m_e c^2)$.
The number density of the target photons, per unit frequency $\nu^\prime$, is $n_\nu(\nu^\prime)$. 
We have considered different contributions to the photon field, namely: 
\begin{enumerate}
\item  The photons resulting from the synchrotron emission
 \begin{equation}\label{eq:IC_nSSC}
	n^\mathrm{SSC}_\nu (\nu^\prime) =\frac{L^\mathrm{SYNC}_\nu (\nu^\prime) }{4\upi R^2}  \frac{\bar{U}}{c h \nu^\prime}  \,,
\end{equation}
with $L^\mathrm{SYNC}_\nu$ the luminosity coming from synchrotron emission and computed as shown in the following eq.~\ref{eq:L},  $R=R_\mathrm{PWN}$ is the region in which the synchrotron emission is produced, and $\bar{U}\simeq 2.24$ is the mean over a spherical volume of the function $U(x)$ defined in \citet{Atoyan:1996}, that accounts for the number density of photons at a given distance considering an isotropic emissivity of the synchrotron radiation in a spherical source. This approximation was also used in this context in \citet{Tanaka2010, Martin2012}.
\item   The photon contribution from the cosmic microwave background (CMB), modeled as a pure black-body
 \begin{equation}\label{eq:IC_nCMB}
	n^\mathrm{CMB}_\nu (\nu^\prime) =\frac{8\upi }{c^3}  \frac{\nu^\prime}{\exp{[h\nu^\prime/(k_B T) -1]}}  \,,
\end{equation}
with $k_B$ the Boltzmann constant and $T_\mathrm{CMB}=2.7$K is the CMB temperature.
\item  The Galactic near-infrared background (N-IR) and the far-infrared (F-IR) one, both modeled as diluted black bodies
 \begin{equation}\label{eq:IC_nIR}
	n^\mathrm{x-IR}_\nu (\nu^\prime) =\frac{15 w_\mathrm{x-IR} h^3}{(\upi k_B T_\mathrm{x-IR})^4}  \frac{1}{\exp{[h\nu^\prime/(k_B T_\mathrm{x-IR}) -1]}}  \,,
\end{equation}
where $w_\mathrm{N-IR} = 5.0$ eV/cm$^3$ and $w_\mathrm{F-IR} = 1.4$ eV/cm$^3$ are the IR energy densities while $T_\mathrm{N-IR} = 3500 $ K and $T_\mathrm{F-IR} = 35$ K the temperatures. 
We have not taken into account the contribution from Bremsstrahlung emission, which is expected to be largely sub-dominant (see, e.g., \cite{Torres2013}).  The  energy densities values used here have been taken from \citet{Torres2014}, not fitted against.  We refer to the former paper for a more detailed accounting of observational data.
\end{enumerate}

\subsubsection{Emissivities and spectrum}

The synchrotron or IC emissivity can then be obtained using 
 \begin{equation}\label{eq:j}
	j^\mathrm{[SYNC,\, IC]}_\nu (\nu, X, Y, Z) = 
	 \int_0^{\epsilon_\mathrm{max}} \mathcal{P}^\mathrm{[SYNC,\, IC]}_\nu (\nu,\epsilon) f(\epsilon) \,d\epsilon  \,,
\end{equation}
where the particle distribution function $f(\epsilon)$ is given by Eq.~\ref{eq:f}.
Integrating this quantity along the line of sight (here assumed to be in the $X$ direction) the surface brightness can be obtained as
\begin{equation}\label{eq:Int}
	I^\mathrm{[SYNC,\, IC]}_\nu (\nu, Y,Z) =  \int_{-\infty}^{\infty} j^\mathrm{[SYNC,\, IC]}_\nu (\nu,\epsilon) \,dX  \,,
\end{equation}
and finally the emitted luminosity is 
 \begin{equation}\label{eq:L}
	L^\mathrm{[SYNC,\, IC]}_\nu (\nu) =  4\upi \int_{V_\mathrm{PWN}} j^\mathrm{[SYNC,\, IC]}_\nu (\nu,\epsilon)\, dX dY dZ \,.
\end{equation}
The integrated flux can be computed form this last expression by weighting properly the luminosity with the source distance $d$: $F_\nu(\nu) = L_\nu(\nu)/(4\upi d^2)$. 

Relativistic effects on the emissivity are properly taken into account, with the corrections to the frequency and emission coefficient in the observer frame given by $\nu_\mathrm{obs}= D\nu$ ad $j_{\nu,\mathrm{obs}}= D^2 j_{\nu}$, with $D=1/[\gamma_w(1-\boldsymbol{\beta}\cdot\mathbf{n})]$ the Doppler boosting factor and $\boldsymbol{\beta}=\boldsymbol{v}/c$ (see e.g. \citealt{Rybicki:1979}).

The HD+B approach to the description of the magnetic field in 2D simulations simultaneously leads to a very good integrated spectrum, as can be seen in Fig.~\ref{fig:spectrum}. 
It is important to recall that in our HD+B model the only free parameters are the initial intensity of the magnetic field ($\eta$), the two spectral indices ($p_L,\, p_H$),  
and the values of the maximum and break energies.
The latter one in particular appears to be very uniform at the end of the evolution, with no significant variation from the injected value ($\epsilon_b=10^5$), due to the very young age of the system. 
%
%
The normalization constant $K_0$ is fixed by Eq.~\ref{eq:sp_norm}.
The pressure scaling is also fixed by the magnetic field (i.e. the initial value of $\eta$) and the age of the system. 
The spectral indices are constrained by fitting the synchrotron integrated spectra. 
What we found is: $p_L=1.10$ for the low energy emitting particles and $p_H=2.55$ for the high energy emitting ones.

The two scalings that we obtain directly by computing the expressions in Eq.~\ref{eq:corr1} and Eq.~\ref{eq:corr2} are: $c_1=0.75$, $c_2=0.14$.
These corrections are mostly due to high energies particles, especially in the present case given the young age of the system. 
Indeed, 
all the particles belonging to the $L$ family have in fact synchrotron ages that are larger than the source age, being then subject to almost no synchrotron losses up to $t_\mathrm{age}$. 
This is not true for the $H$ family of particles, which is subject to strong energy losses from the beginning, especially for those particles emitting at $\nu \gtrsim 10^{12}$ Hz.
This might be different for older sources.

The high-energy IC spectrum is finally computed using the values of the parameters constrained by the fit of the synchrotron components, without introducing ad hoc normalization constants  to fit the gamma-ray data.
Note that within the HD+B model, we were not forced to disentangle the two emission mechanisms with the use of different normalizations nor to artificially steepen the high-energy synchrotron spectrum, as it was usually necessary with 2D MHD models in order to overcome the lack of energy losses caused by the wrong magnetic field \citep{Volpi2008, Olmi:2014}. 
For the same reason, the latter models show an overestimated gamma-ray spectrum. 
On the contrary, we are able to reproduce also the correct ratio between the IC and synchrotron spectra.

\begin{figure}
        \includegraphics[width=0.45\textwidth]{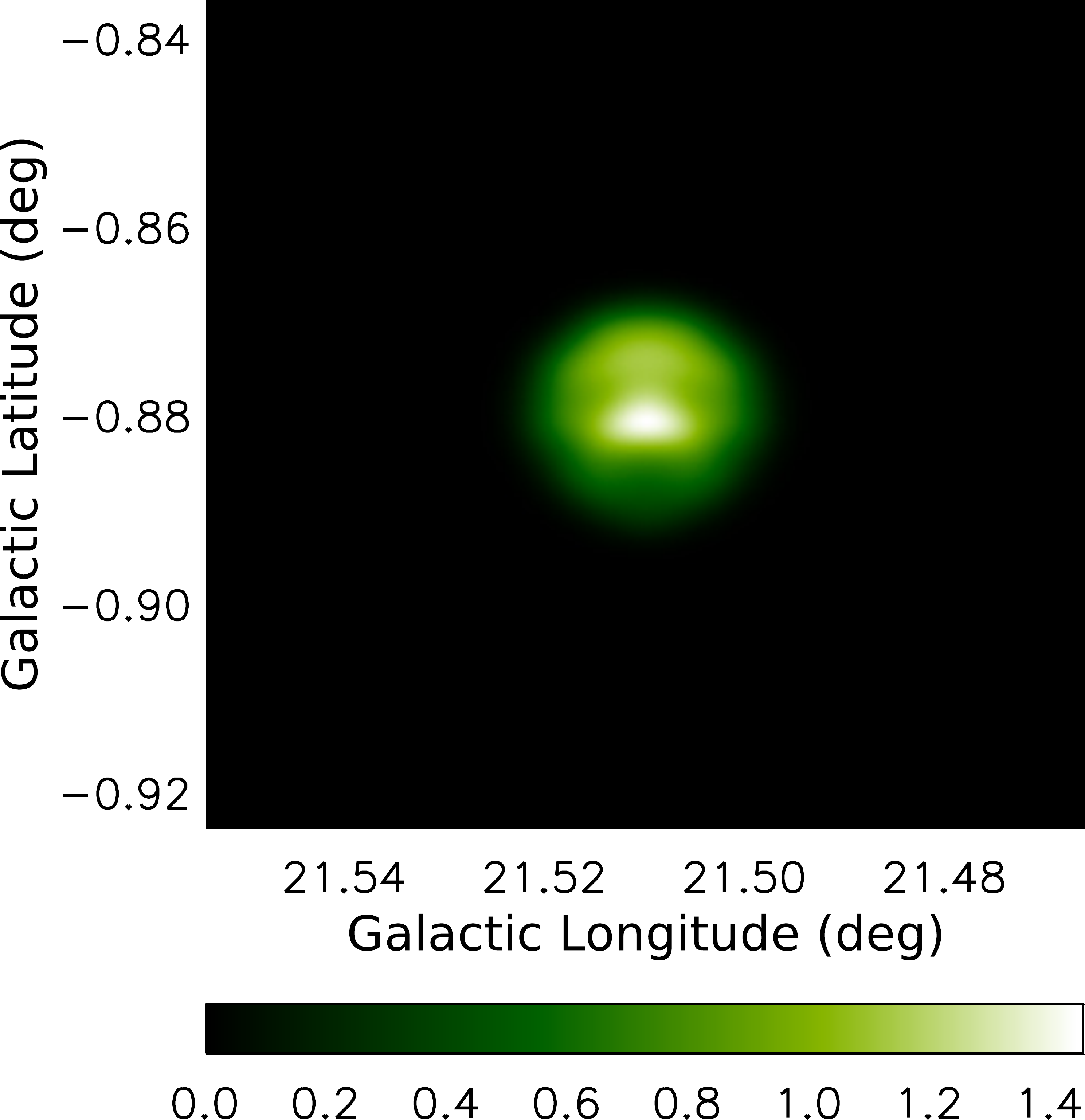}\\
	\caption{Radio surface brightness map at $\nu=1.4$ GHz. The intensity is given in units of mJy for a direct comparison with available data and it is expressed by the color scale.}
	\label{fig:radio}
\end{figure}

We show a surface brightness map of the radio emission at 1.4 GHz in Fig.~\ref{fig:radio}.
The original high resolution synthetic map has been elaborated to allow a direct comparison with observations, following the data presented in \citet{Bietenholz:2011} and obtained with the National Radio Astronomy Observatory Very Large Telescope (NRAO VLA). 
The  original map we obtained was thus convolved with a gaussian Point Spread Function (PSF) having a full-width-half-maximum FWHM of $[18.8\arcsec,\,13.8\arcsec]$. 
We noticed that a higher resolution radio image was presented in \citet{Bietenholz:2008}, but the emission there is strongly dominated by the thermal filaments, which were not subtracted from the map, making it not a good target to compare with synthetic maps that do not include this additional component.
%
%
We found a  total flux at 1.4 GHz of $7.36$ Jy, in perfect agreement with the $7.0 \pm 0.4$ Jy deduced from the NRAO VLA data in the cited work. 
The authors  also found a peak surface brightness of $0.63\pm0.03$ Jy/beam which corresponds to the maximum intensity of our original map ($0.61$ Jy/beam).
This value appears to be consistently lower to $0.43$ Jy/beam in the convolved map.

\begin{figure*}
        \includegraphics[width=0.99\textwidth]{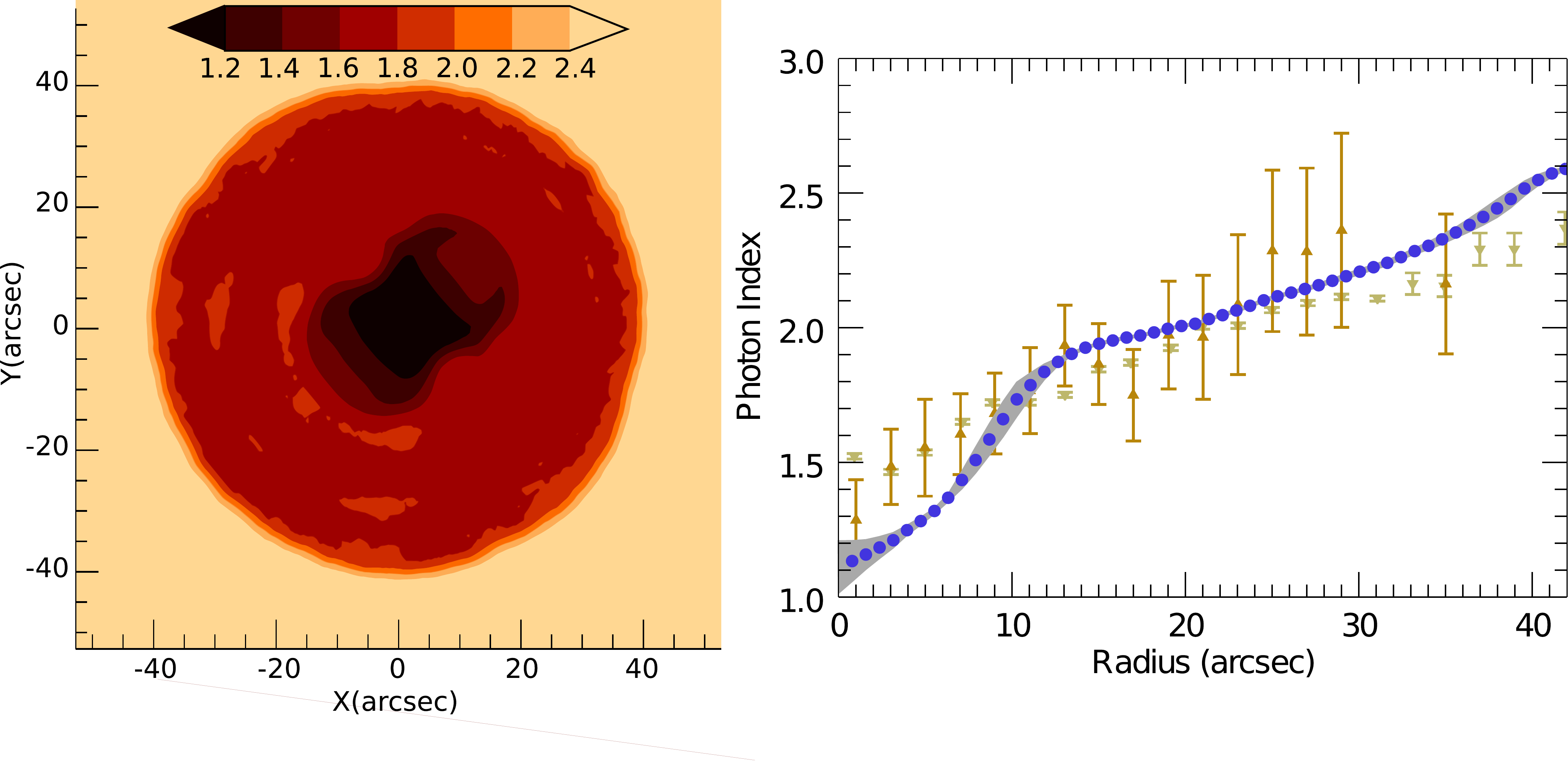}
	\caption{%
	\emph{Left-hand panel:}  2D map of the photon index for a particular inclination and rotation of the PWN.
	\emph{Right-hand panel:} Plot of the radial profile of the X-ray photon index in the range of 0.5-8 keV, computed as described in the main text: blue circles show the photon index averaged over the inclination angle while the gray area is the standard deviation. Points with the triangle symbols (up and down) are taken from Fig.6-7 \citet{Guest:2019}, together with their corresponding errors, representing the PWN photon index with the background subtracted (down triangles) and the photon index of the entire remnant. 
	}
	\label{fig:Xspind}
\end{figure*}

\subsubsection{Spatial-distribution of the X-ray photon index}

We show the results on the X-ray photon index variation in the energy range $0.5-8.0$ keV in Fig.~\ref{fig:Xspind}, compared with recent results presented by \citet{Guest:2019}. 
The photon index has been computed from the spectral slope $p=-\log{[I_\nu(\nu_2,\,Y,\,Z) \, I_\nu^{-1}(\nu_1,\,Y,\,Z)}/\log{(\nu_2 \nu_1^{-1})}$, with $I_\nu(\nu,\,Y,\,Z)$ being the synthetic intensity as given by eq.~\ref{eq:Int}. 
With this procedure we have built the $(Y,\,Z)$ maps of the photon index in the considered energy range, using 10 steps in frequency for different inclination angles of the PWN with respect to the line of sight.
A representative case is shown in the left-hand panel of the figure, for an inclination angle with the line of sight of $\sim 60^\circ$ and a rotation of $\sim 45^\circ$.
We then extract for each map at a fixed inclination the average profile, by averaging along different shells centered on the pulsar, similarly to the procedure used for the real data.
The final profile, shown in the right-hand panel of Fig.~\ref{fig:Xspind} is then obtained averaging on all the different profiles extracted from the maps at the various inclinations.
Since no constraints are currently available on the real inclination of the source we also show the standard deviation computed on all the different possible inclinations.
In any case, as Fig.~\ref{fig:Xspind} shows, the inclination effect is negligible in the first place and averaging is used to improve statistics. 
This may not be the general case, but results here from the spherical symmetry of the source. 
In general, a measurement of the profile could in principle be used to choose a range of inclinations that most closely reproduce it.
Alternatively, a measurement of the profile could in principle be used to choose a range of inclinations that most closely reproduce it.

The radial variation of the X-ray spectral index within the nebula radius ($\sim 42\arcsec$) shows a remarkable agreement with predictions from the data analysis, despite the limitations of our model in accounting for the X-ray inner nebula properties.
It shows a harder profile near the PWN center that becomes softer and softer when moving outwards. 
The region with radius $\gtrsim 40\arcsec$ cannot be compared directly since it lies outside the compact PWN, which is the part we are simulating here. 
For this reason we simply stop our prediction at $\gtrsim 40\arcsec$, noticing that for $r>0.42\arcsec$ observations predict a photon index of $\sim2.6$ (see for a comparison Fig.~7 of \citealt{Guest:2019}).

\subsubsection{TeV emission map}

\begin{figure*}
        \includegraphics[width=\textwidth]{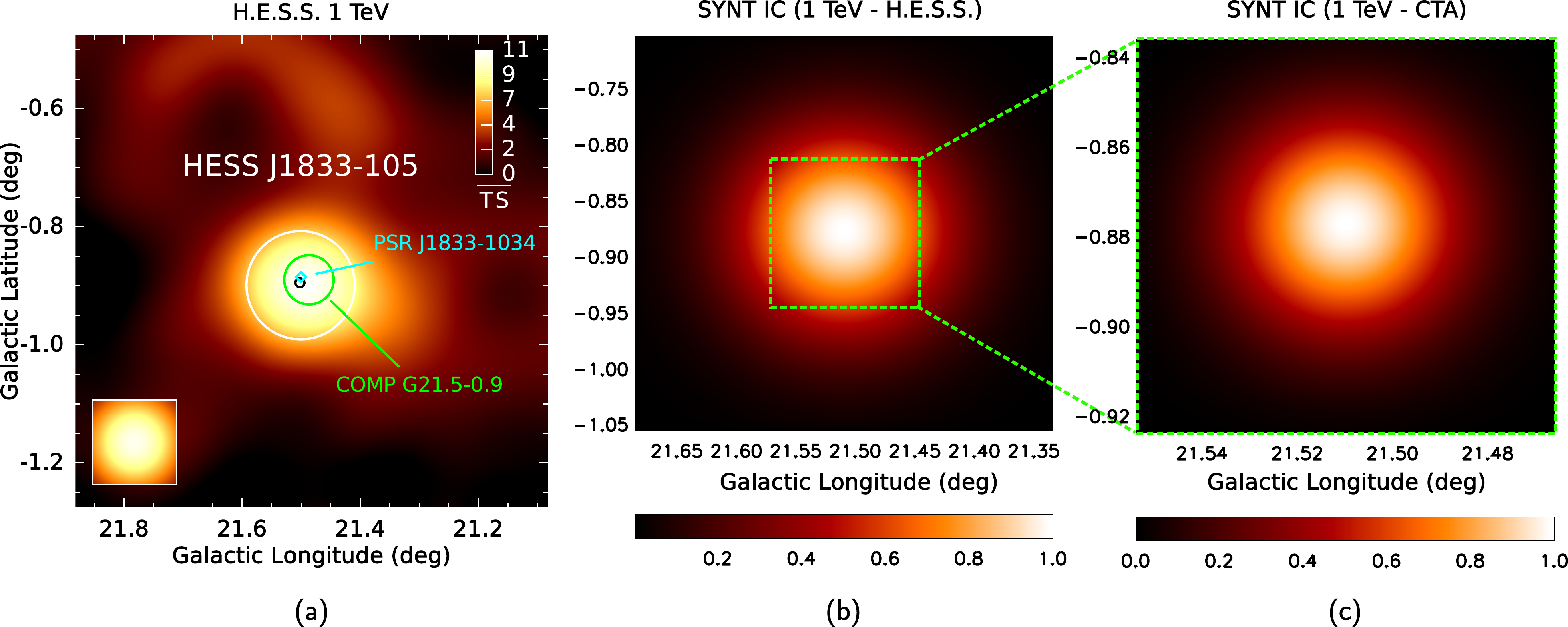}\;
	\caption{Qualitative comparison of the IC emission at 1 TeV from data and simulations. From left to right: (a) H.E.S.S. significance map of the entire J1833-105 remnant, from which G21.5-0.9  (highlighted with the green circle) is considered to produced the radiation. This image taken from the H.E.S.S. GPS paper \citealt{H.E.S.S.-Collaboration:2018} and available at \url{https://www.mpi-hd.mpg.de/hfm/HESS/hgps/}); (b) synthetic surface brightness map of the IC emission smoothed with the H.E.S.S. PSF of $0.16^{\circ}$; (c) synthetic surface brightness map of the IC emission smoothed with the expected CTA PSF of $ \sim 0.16^{\circ}/4$.
	In all cases the spatial extents are given in Galactic coordinates, longitude and latitude in degrees, but note that in panel (c) the range is different to zoom in the source with respect to the H.E.S.S. maps. The surface brightness maps computed are in cgs units using the normalization fixed by the fitting of the integrated spectrum and they are here scaled by the maximum value of the intensity.}
	\label{fig:ICmap_CFR}
\end{figure*}

We finally consider the gamma-ray IC emission, compared with available observations and possible future ones.
The H.E.S.S. view at 1 TeV of G21.5-0.9 is shown in the leftmost panel (a) of Fig.~\ref{fig:ICmap_CFR}. 
Here, the instrumental resolution is shown in the bottom-left square of the image, corresponding to a PSF with $0.16^{\circ}$ FWHM. 
The observed image is given in terms of the measured significance, corresponding to the over-imposed color scale. 
The PWN only represent a small inner part of the emission map (within the green circle), the available resolution does not allow for spatially resolving it.
We show the IC emission at 1 TeV as obtained from our simulations, after being convolved with the same PSF of H.E.S.S., 
in the central panel of the same Fig.~\ref{fig:ICmap_CFR}. 
The map is normalized to its maximum value for an easier qualitative comparison with the significance map of panel (a). 
As expected, since the angular resolution of the instrument is larger than the extension of the PWN itself ($\sim 0.024^\circ$ in diameter), the convolution of  the original image with the instrumental PSF produces a complete loss of morphological information. 
With the H.E.S.S. resolution the compact nebula is in fact seen as a point source, with its morphology starting to be resolved for a PSF$\lesssim$PSF$_\mathrm{H.E.S.S.}/6$.
A quantitative comparison with resolved IC emission is then not possible since the estimated range of fluxes as obtained in the analysis of the H.E.S.S. GPS take into account the contribution from the whole composite source, with no possibility of isolating the contribution from the PWN alone. 
In panel (c) of Fig.~\ref{fig:ICmap_CFR}, we finally show the predicted gamma-ray image of G21.5-0.9 as it could be observed with CTA, using updated estimations for the angular resolution in the $0.8-1.25$ TeV energy range of a FWHM $\sim0.04^\circ$, as recently used by \citep{Mestre:2020}. 
With this resolution the PWN is better resolved, and this opens the possibility of morphology matching studies at high energies --even if subsequent improvements in resolution will be needed to really see the details of the inner nebula \citep{Mestre:2020}.

\section{Summary and Conclusions}
\label{sec:conclusions}
\begin{figure*}
        \includegraphics[width=\textwidth]{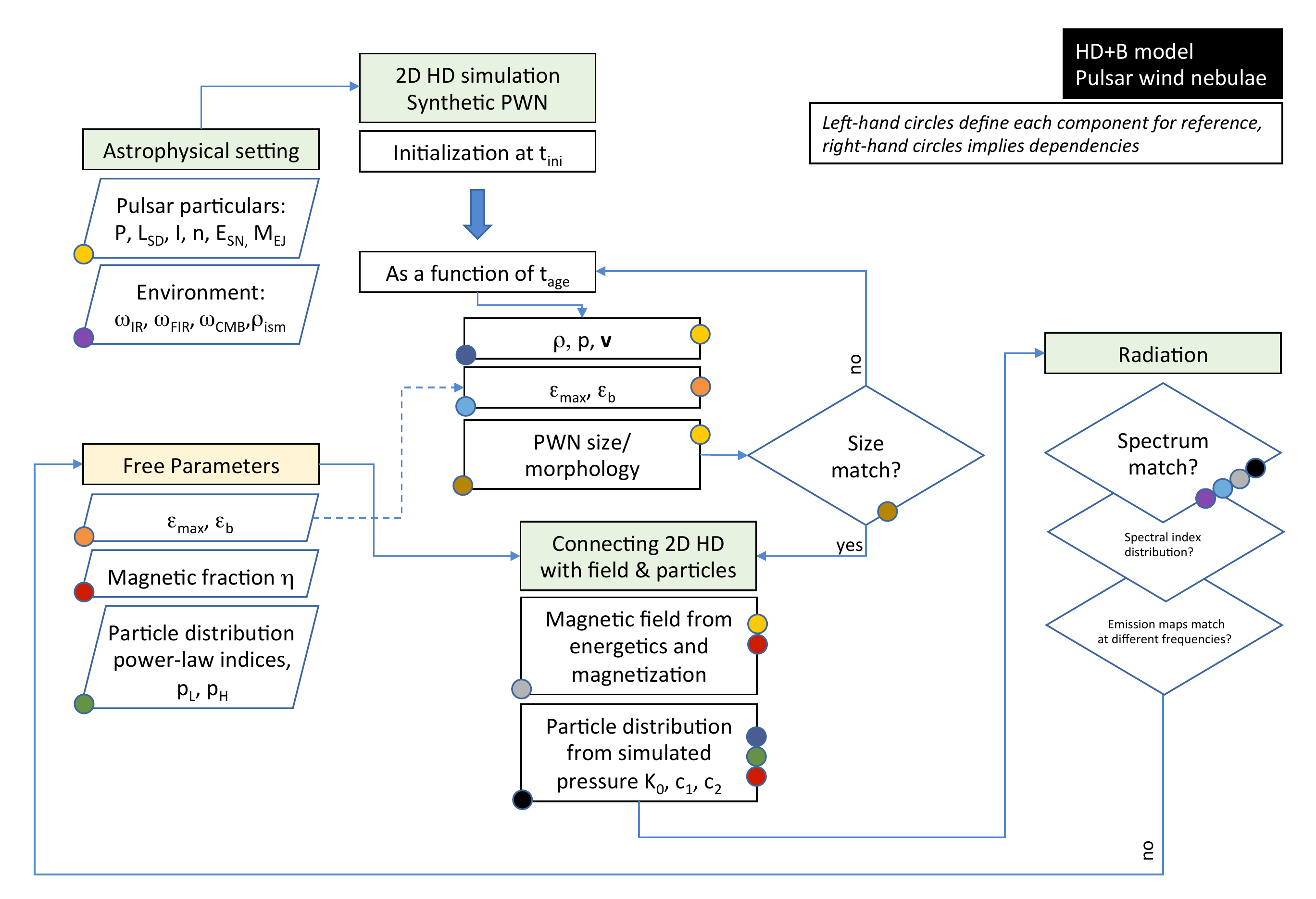}\;
	\caption{Conceptual flowchart of the HD+B model, with the logical path and dependency on variables/quantities around the different iterations of the model.
	The dashed line is used to show that the evolution of these parameters are done via tracers.
	 Colored circles in the left-hand part of each component define them for subsequent reference in the plot, whereas right-hand circles implies dependencies. This is used to reduce the number of connecting arrows, yet  intuitively conveying the general setup of the model.}
	\label{fig:HD+Bscheme}
\end{figure*}

In this paper, we have presented a novel approach for the modelling of PWNe. 
Our hybrid scheme connects relativistic HD numerical simulations with radiative models. 
The main motivation for using such a hybrid approach comes, on the one hand, from the evidence that 2D MHD models are strongly affected by their geometrical limitations. 
They were shown to be non-reliable for producing large scale predictions of  PWN's emitting properties, being on the contrary very successful in reproducing the inner nebula characteristics, see e.g., \cite{Porth:2014}.
On the other hand, whereas 3D MHD simulations were proved to be the solution to the limitation of 2D models, with a very impressive capability of accounting for both small and large scale properties of those sources,
they are extremely expensive in terms of numerical resources and time costs, requiring months of continuous operations and millions of cpu hours to reproduce just a few hundreds of years of evolution of single PWN \citep{Olmi:2016}.
The latter precludes developing identification and characterization methods of unknown PWN based on 3D MHD simulations, at least for now. 
In such a case, in addition, not only the age of the system (which may or may not be known) would be a problem, but the much larger number of free parameters that one would need to range to find a fitting region of the enlarged phase space makes this alternative simply beyond reach.

Here we proposed a way out to this problem. 
The limitations of the 2D approach can be relaxed if the magnetic field is detached from the dynamics. 
In this case, the dynamical evolution of the system is not affected by the field distortion.
But we propose to do this decoupling in a such a way that the field gets linked with the dynamical evolution, under simple, and physically-motivated assumptions; this is the central aspect of our HD+B model.
This allows us to have a field which is time and spatially dependent in the PWN, from which an emission spectrum can be derived, at the same time that we can use the HD simulations directly to define the particle population and the morphology.

The way in which we treat the magnetic field is the key difference with former approaches 
in which the magnetic field is evolved as part of the dynamics (\citealt{Del-Zanna:2006} and subsequent papers of the sort).
On the contrary, in our HD+B method the magnetic field is introduced in analogy with what is done in pure radiative models. 
The field is directly linked to the pressure and has no back-reaction on the dynamics of the PWN.
It is not affecting the evolution of the system as it happened with previous 2D MHD models.
In these latter models, the resulting emission maps reflect the artificial distribution and intensity of the field introduced by the dimensionality, being not able to reproduce the morphology and spectra of the sources well. 
The field appears always to be too high at the center but too low at the boundaries in 2D MHD simulations, strongly affecting the resulting emission.
Since the magnetic field has such a strong impact on the overall emission at the different frequencies, those models cannot account for a single normalization of the synchrotron particle distribution function and require to disentangle the normalization (hence introducing an extra parameter) of the two families ($L$ and $H$, for low and high-energy particles, respectively), as well as to change by hand the $H$ spectral index to reduce the effect of the underestimated synchrotron losses.
This is all avoided here.
In our model, the distribution of the magnetic field is quasi-uniform, as is the pressure, as expected from 3D models on large scales. 
The average value of the field is the one expected in the total volume of the nebula, giving thus the correct broad-band spectrum, with no need of keeping disentangled the normalizations of the distribution functions nor of modifying artificially the spectral indices.

We also note that the normalization of the pair spectrum is done in a natural way, equating its pressure to the thermal pressure for a given set of assumed spectral indices (that are free parameters for fitting the spectral yield).  
We compute the thermal pressure also from the simulation, considering the total pressure (simulation) result, but subtracting from it the amount of energy (pressure) that powered the magnetic field along
history, and the amount of particles that died along the evolution due to losses.
This is done in a time-dependent way, for each time of interest $t_{age}$, so that the correction factors are completely fixed numerically computing integrals describing the former energy/pressure sinks. 
The only free parameters of our HD+B approach to determine the particle distribution are then the spectral indices and the break energy, that of course must be assumed comparing with available observations. 

As a summary, 
Fig.~\ref{fig:HD+Bscheme} shows a conceptual flowchart of the HD+B method, highlighting the different phases of the model and how these depends on the free parameters.

We have also presented here the first application of this new HD+B method, using the case of the young and well-characterized PWN G21-5-0.9.
The simulation was initialized based on observational data, where available, and earlier results from pure radiative models in the other cases.
Using our synthetic PWN results, we were able to compute a wide range of multi-wavelength results which were found to be in good agreement
with observations.
In particular, we presented a self-consistent way to compute the PWN entire electromagnetic spectrum, from radio to gamma-rays, based on the few free parameters of the HD+B model,
among them, the spectral indices of the particle distribution function, considered as usual a power law in energy with different indices for the low energy and high energy components.
We got a very good fit to the spectrum, especially considering the lack of degree of freedom in the fitting procedure (for instance, the normalization is fully constrained given 
the indices and the HD simulation).
Once the fitting parameters were fixed, we have also computed emission maps at radio and gamma-ray energies.
We compared the radio map at 1.4 GHz with the data presented in \citep{Bietenholz:2008}, obtaining a remarkable agreement with the expected total flux and maximum intensity. 
The 1 TeV gamma-ray surface brightness map was compared with available data from the H.E.S.S. Galactic plane survey and 
we have also verified that the CTA view on the source will allow for a deeper analysis, being the source spatially resolved.

The limitation of the proposed method is found in its inability of accounting --in the general case-- for fine structures usually observed at X-rays and connected to the inner geometry of the magnetic field.
This method was not devised for modelling the inner nebulae, so that a morphological comparison in the inner X-ray scale is not advisable.
The X-rays are in fact mostly shaped by the geometry of the magnetic field and velocity flux in the inner nebula (responsible for the jet-torus shape observed in Crab and other PWN), 
that cannot be reproduced with a quasi-uniform magnetic field.
In the case of the target considered for this first application, moreover, the X-ray emission is by itself very puzzling, with a spatial extent greater than the radio one. 
This might indicate a diffusion-dominated nature of the X-ray emission near the PWN boundary, which treatment is beyond our intents.
However, as we have explicitly shown, the HD+B can still be constrained by X-ray observations, being able to compute and compare  the expected variation of the photon index in the keV band.
Also here, we have found a remarkable agreement between model and data.

The development of this new method is motivated by the challenge of interpreting the large amount of existing and forthcoming PWN detections, to be brought by new facilities in radio (SKA and pathfinders) and gamma-rays (CTA). 
For the latter, PWNe are expected to be the dominant sources, with hundreds of new detections foreseen, and an almost complete coverage in the near Galaxy.
We focused in developing a model able to deal with such a large number of detections, 
being especially suited not just to describe PWN of known pulsars, but to make inferences from detections in cases where the system is much less known as the one treated here. 
Further research in this direction is proceeding. 
We foresee two obvious future applications of the presented model. 
On the one hand, we intend to investigate the possible degeneracy of the predicted properties of the source against variation of the physical parameters that are not constrained by observations. 
On the other hand, we will apply the same method to try and model other sources, in particular considering more evolved systems, for which the interaction with the surroundings may be of larger relevance.

\section*{Acknowledgements}
%
We acknowledge the ``Accordo Quadro INAF-CINECA (2017-2019)''  for the availability of high performance computing resources and support. Simulations have been performed as part of the class-B project ``Morphology matching of Pulsar Wind Nebulae produced with HD numerical simulations
'' (Olmi \& Torres). 
We acknowledge Elena Amato, Rino Bandiera, Niccol\`o  Bucciantini, and Jonatan Martin for fruitful discussions. 
This work has been supported by grants PGC2018-095512-B-I00, SGR2017-1383, AYA2017-92402-EXP, 2017-14-H.O ASI-INAF and INAF MAINSTREAM.

\footnotesize{
\bibliographystyle{mn2e}
\bibliography{olmi.bib}
}

\appendix

\section{Comparing different resolutions in 2D HD}\label{sec:a1}
\begin{figure*}
        \includegraphics[width=0.95\textwidth]{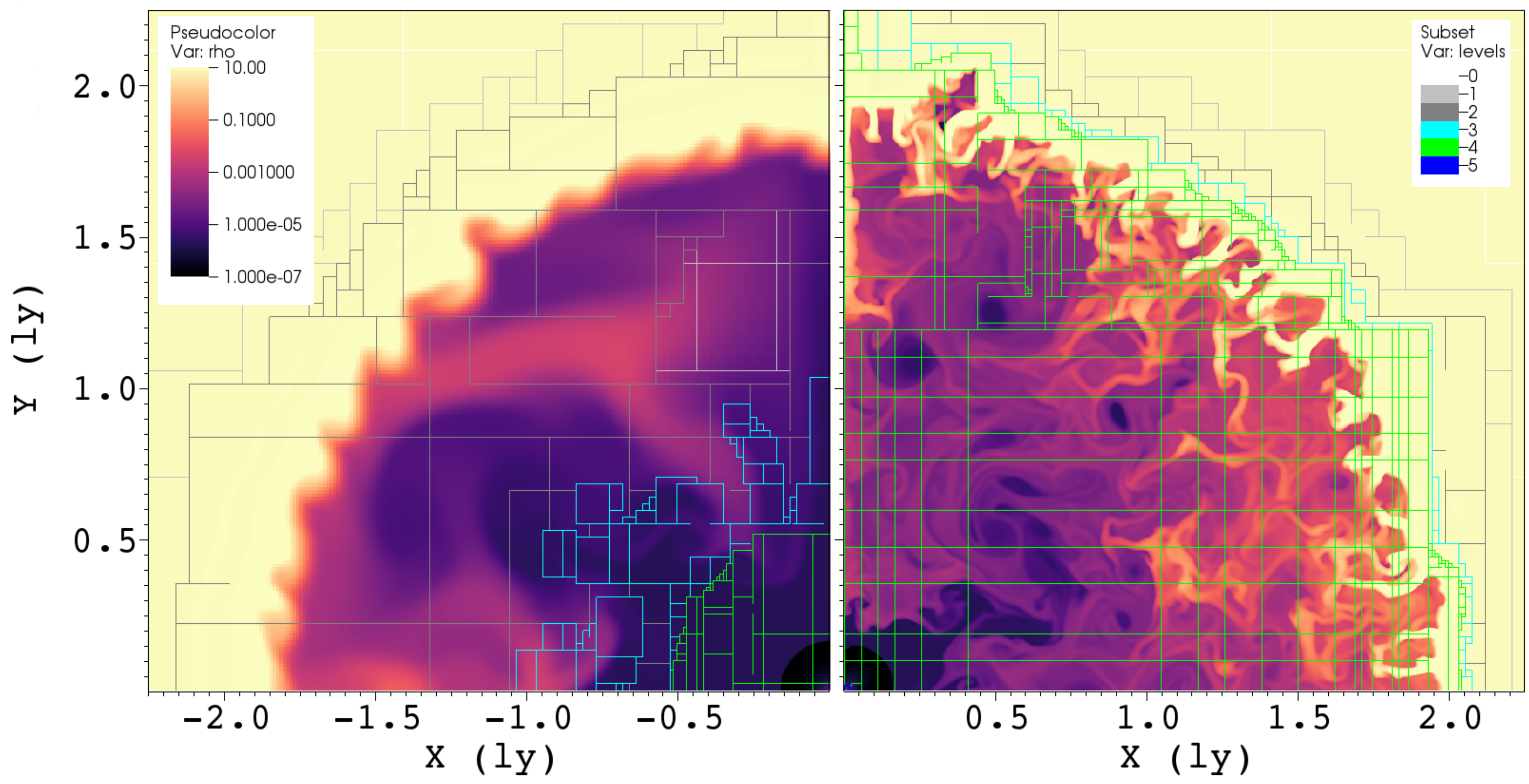}\;
	\caption{Plots of the density at $t=450$ yr, in logarithmic scale and units of 1 particle$/$cm$^3$, with the color coded bar shown. 
	\emph{Left panel:} run with  $[272\times544]$  base grid and 4 AMR levels (shown as colored lines, with legend shown).
	\emph{Right panel:} run with  $[272\times544]$  base grid and 5 AMR levels and with no restriction for the increasing zones of finest levels. Please notice that the 5th level (blue color) is located at the center of the grid, resolving only the injection region.
	The images were produced with the VisIt \citep{Childs:2012} open-source analysis tool. 
	See the text for a discussion.}
	\label{fig:vort_res}
\end{figure*}

In Fig.~\ref{fig:vort_res} we show a comparison of the same dynamical quantity (the density) at $t=450$ yr with different resolutions of the grid.
AMR blocks of different levels are also shown as contours of different colors. 
%
This comparison may help in clarifying the point we made in Sec.~\ref{subsec:details_setup}, when discussing how to define the requested resolution of the simulation.

We show the simulation used for the analyses presented in this paper, with a base grid of $[272\times544]$ cells plus four AMR levels (panel on the left), with a second run with the same base grid but five refinement levels (panel on the right).
Moreover the two panels differ on the spatial definition of the subsequent increasing resolution levels, as can be easily seen if comparing the same AMR level (color) in the two figures. 
For example, in our reference simulation, the fourth level was only activated in the inner nebula ($r\lesssim 0.5$ ly), whereas in the other case it is active in the entire PWN.
This of course requires a longer time to run the same simulation.
Moreover it also has strong effects on the behavior of the RT instability at the PWN contact discontinuity, making the vortexes smaller and smaller. 
The inner material is then strongly mixed up with the one coming from the ejecta, percolating towards the center due to the RT fingers. 
This strong development of turbulent mixing at smaller and smaller scales of the outer and inner densities will end up in the consumption of the PWN contact discontinuity,  making the code unstable.
The strong correlation of the RT instability efficiency in mixing up the nebular material with the outer one was already pointed out in \citealt{Blondin:2001}, where fingers were seen to penetrate deeply the nebula.
We also found that when considering very high resolutions at the border, the mixing become very efficient. Moreover the RT fingers become able to penetrate very deeply the PWN, eventually down to the center.
The same effect was not observed in the MHD framework, where the RT fingers develop in a finite volume of the nebula, with the longest ones penetrating for $\sim1/4$ of the nebula radius, as expected from observations \citep{Porth:2016}. 
We think that the presence of the magnetic field, and thus the magnetic pressure, contribute to sustain the system against mixing, and that the very efficient penetration we observe in our high-resolution runs is purely an effect of the HD scheme.
We then keep the resolution low enough not to allow the RT fingers to reach the center of the nebula, being then sure that the PWN volume remains well identified by our numerical tracers. 
We notice in any case that the underestimation of the complexity of the RT structure at the CD is not going to affect our results, since we do not compare with high-resolution images, given the intrinsic limitations of our approach in that sense.
On the other hand,  the resolution in the inner nebula must be high enough in order to allow for the freshly injected material at the TS to diffuse and be mixed with the older one already there.
This can be ensured if the dimension of the smallest scales in the inner nebula are $\lesssim r_\mathrm{TS}$.
When defining the proper resolution of an HD simulation both competing facts must be then properly considered.
%


\end{document}